\documentclass[runningheads]{svmult}

\usepackage{makeidx}   
\usepackage{graphicx}  
\usepackage{subeqnar}  
\usepackage{multicol}  
\usepackage{physprbb}  
\makeindex             
\begin{document}
\title*{Freeze-out Effects in Hydrogen and Helium Lines\protect\newline 
        of SN 1987A at the Early Photospheric Epoch} 
\toctitle{Freeze-out Effects in Hydrogen and Helium Lines\protect\newline 
          of SN 1987A at the Early Photospheric Epoch} 
%
%
\titlerunning{Freeze-out Effects in Hydrogen and Helium Lines of SN 1987A}
%
\author{Victor~P.~Utrobin\inst{1,2}
\and Nikolai~N.~Chugai\inst{3}}
\authorrunning{V.~P.~Utrobin and N.~N.~Chugai}
%
%
\institute{Institute of Theoretical and Experimental Physics, 117259 Moscow, Russia
\and European Southern Observatory, D-85748 Garching, Germany
\and Institute of Astronomy of Russian Academy of Sciences, 109017 Moscow, Russia}

\maketitle

\begin{abstract}
 We have developed a time-dependent model of ionization, excitation  and 
    energy balance of SN~1987A atmosphere at the photospheric epoch 
    to study early behavior of hydrogen and helium lines. 
 The ionization freeze-out effects play a key role 
    in producing both the strong H$\alpha$ during nearly the first month and  
    the He~I 5876\,\AA\ scattering line on day 1.76.
 Using an extended reaction network between hydrogen molecules 
    and their ions demonstrates that ion-molecular processes are likely 
    responsible for the blue peak in H${\alpha}$ profile at the Bochum 
    event epoch.
\end{abstract}

\section{Introduction}
 We understand a spectrum formation of SNe~II not better than 
    we do the hydrogen and helium spectra of SN~1987A at the early 
    photospheric epoch.
 Despite many attempts to explain H$\alpha$ and He I 5876\,\AA\ lines
    we are still far from satisfactory results~\cite{viu01,viu02,viu03,viu04}.
 The stumbling block is that any model produces too week H$\alpha$ and 
    He I 5876\,\AA\ at the early photospheric epoch. 

 Yet, it has already become clear that a residual ionization (freeze-out effects) 
    may result in the enhanced excitation of hydrogen compared to steady-state 
    regime~\cite{viu05,viu06}.
 However, in our previous study~\cite{viu06} we assumed simple
    laws for the  behavior of electron temperature in the atmosphere.
 Here we will confirm a vital role of freeze-out effects in the hydrogen 
    excitation at the photospheric epoch using the upgraded model in which 
    time-dependent chemical kinetics is solved along with time-dependent energy 
    balance.
 Moreover, here we consider neutral helium, not only hydrogen, and will show 
    a key role of the time-dependent kinetics for He I 5876\,\AA\ line too.

 We also recapitulate here our previous results on modeling the blue peak of 
    H$\alpha$ at the Bochum event phase upon the basis of the time-dependent 
    kinetics with hydrogen-composed species~\cite{viu06}. 
 This will demonstrate that not only first-order effects, like freeze-out, 
    but subtle molecular processes may be also important for hydrogen spectrum 
    formation in SN~II.

\begin{figure}[t]
\begin{center}
\includegraphics[width=0.9\textwidth]{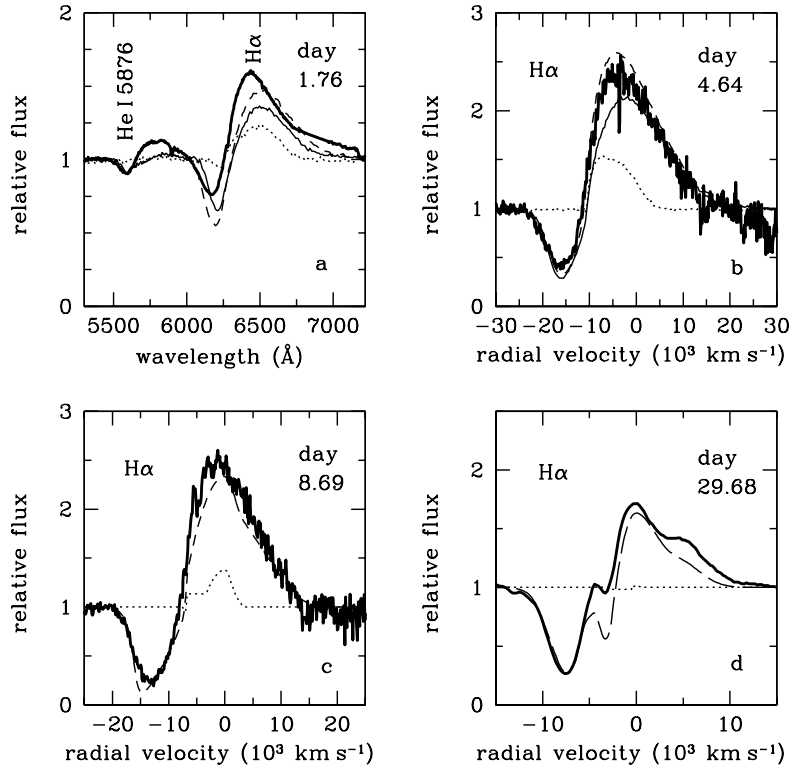}
\end{center}
\caption[]{
 Time evolution of H$\alpha$ from day 1.76 to 29.68 (\textbf{a-d}) and 
    He I 5876\,\AA\ on day 1.76 (\textbf{a}) in SN~1987A.
 The observed spectra from day 1.76 to 8.69~\cite{viu12} and on day 29.68~\cite{viu13}
    ({\it thick solid line\,}) are compared to those computed 
    with the full time-dependent energy balance ({\it thin solid line\,}), 
    with adiabatic evolution of the kinetic temperature ({\it short dashed line\,}), 
    with the electron temperature equal to the local radiative temperature 
    ({\it long dashed line\,}), 
    and for the steady state ({\it dotted line\,}) 
}
\label{utrobinF1}
\end{figure}

\begin{figure}[t]
\begin{center}
\includegraphics[width=0.9\textwidth]{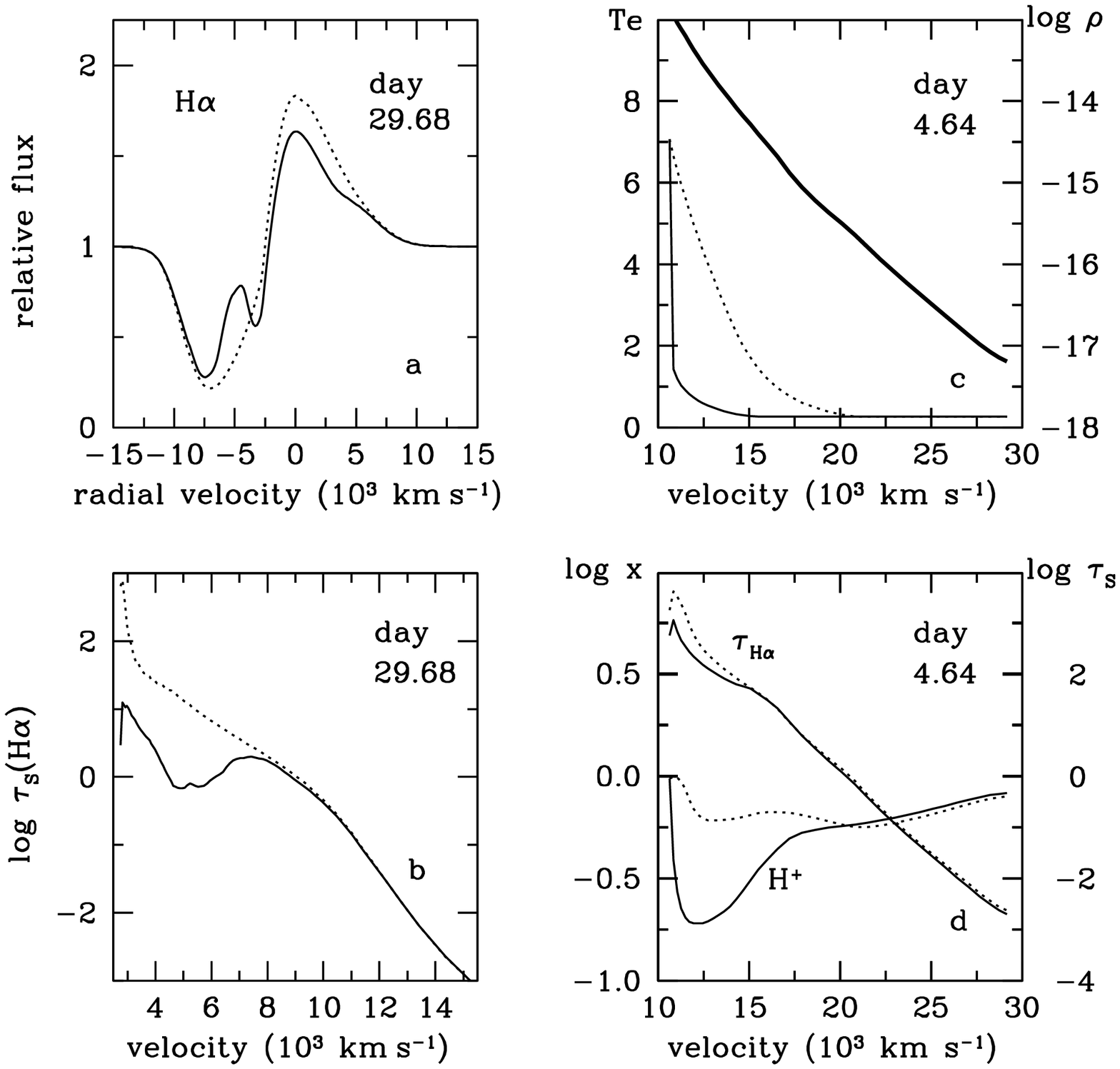}
\end{center}
\caption[]{
 {\it Left panel}: The role of molecular reactions for the Bochum event. 
 (\textbf{a}) H$\alpha$ profiles on day 29.68 calculated 
     in time-dependent approach with ({\it thin solid line\,}) and without 
     ({\it dotted line\,}) molecular reactions. 
 (\textbf{b}) The behavior of the Sobolev optical depth of H$\alpha$ in both models. 
 {\it Right panel}: Physical conditions in the supernova atmosphere ($v>10700$\,km/s) 
    on day 4.64.
 (\textbf{c}) The density ({\it thick solid line\,}) and electron temperature 
    (in units of 1000\,K). 
 The temperature is given for the model with the full energy balance 
    ({\it thin solid line\,}) and for that with the adiabatic approximation
    ({\it dotted line\,}). 
 (\textbf{d}) The behavior of the Sobolev optical depth for H$\alpha$
    and fractional abundance of H$^+$ for both models 
}
\label{utrobinF2}
\end{figure}

\section{Model and Input Physics}
 The atmosphere model is based on the hydrodynamic model~\cite{viu07} with
   a 15$M_{\odot}$ ejecta and a kinetic energy of 1.9$\times$10$^{51}$\,erg.
 The radiation field is treated in the approximation of sharp photosphere.
 For t$<$1.8 days the radiation field in continuum is described by the photospheric
    radius and effective temperature, taken from the hydrodynamic model, and 
    at later epoch by the empirical values~\cite{viu08} and 
    by the UV and optical observations~\cite{viu09}.
 The line radiation transfer is treated in the Sobolev approximation~
    \cite{viu10,viu11}.
 Electron temperature in the atmosphere is determined from solving the time-dependent
     energy equation.

 The following elements and molecules are calculated in non-LTE chemical kinetics:
    H, He, C, N, O, Ne, Na, Mg, Si, S, Ar, Ca, Fe, H$^{-}$, H$_{2}$, H$_{2}^{+}$,
    and H$_{3}^{+}$.
 All elements but H are treated with the three ionization stages.
 The level populations of H and He I are calculated for 15 and 29 levels,
    respectively, while other atoms and ions are assumed to consist of
    the ground state and continuum.
 The reaction network involves all bound-bound and bound-free, radiative and
    collisional processes for atoms and ions, and 7 radiative and 37 collisional
    processes for molecules.

\section{Results and Discussion}
 We have developed the time-dependent chemistry model coupled with the 
    time-dependent energy balance at photospheric phase and investigated spectra 
    of SN 1987A.
 The fit between the observed and calculated H$\alpha$ profiles on days 1.76, 4.64,
    8.69, and 29.68 and those of He I 5876\,\AA\ line on day 1.76 is fairly good 
    in Fig.~1 indicating that our model reflects the reality reasonably well.
 However, we admit that on day 1.76 our assumption on the sharp photosphere
    may be responsible for the deficit of the calculated H$\alpha$ and 
    He I 5876\,\AA\ net radiation. 

 The time-dependent effects in the H$\alpha$ and He I 5876\,\AA\ lines are remarkable
    in producing the strong absorption both in H$\alpha$ and He I 5876\,\AA\ lines 
    on day 1.76 (Fig.~1a).
 Moreover, a relative contribution of time-dependent effects to H$\alpha$ 
    increases in time (Fig.~1). 
 No doubts, the ionization freeze-out plays a key role in producing the
    ionization and excitation of hydrogen during at least first month.
 At later epoch the non-thermal ionization and excitation from radioactive 
    $^{56}$Ni decays become important~\cite{viu07}.

 After about day 20 molecular processes neutralizing ionized hydrogen (mainly 
    H$^{-}$ + H$^{+}$ $\rightarrow$ 2\,H) dominate over recombination close to 
    the photosphere.
 This results in the non-monotonic radial dependence of the H$\alpha$ Sobolev optical 
    depth that is blamed for the blue peak in H$\alpha$ at Bochum event phase
   (Figs.~2a,b).
 Earlier we have used the adiabatic approximation for electron temperature~\cite{viu06}.
 While it is a good approximation for the outer layers, we demonstrate that the full 
    time-dependent energy balance should be solved to produce more confident results 
    in the time-dependent models (Figs.~2c,d).

\section{Conclusions}
 Our present study of the time-dependent effects and the influence of molecules and 
    radioactivity in the photospheric epoch of SN~1987A may be summarized as follows:
\begin{itemize}
\item the time-dependent effects are a key prerequisite for the strong H$\alpha$ 
      line during nearly the first month and the He I 5876\,\AA\ line on day 1.76;
\item the additional excitation from radioactive $^{56}$Ni decays is not required 
      to account for the H$\alpha$ line during about the first month;
\item the molecular processes among hydrogen-composed species turn out important 
      factor of hydrogen neutralization which is manifested by the emergence of 
      the blue peak of H$\alpha$ at the Bochum event phase.
\end{itemize}
 We believe that these effects are important for normal SNe II-P at the 
    plateau epoch as well.

\section*{Acknowledgments}
 V.P.U. is grateful to F. Patat and B. Leibundgut for hospitality at the ESO
    and the financial support.
 This work is partially supported by the RFBR (project 01-02-16295).

\end{document}